\documentclass[10pt,twocolumn,amssymb,amsmath,aps,pra]{revtex4-1}
\usepackage{graphicx}

\begin{document}

\title{Noiseless amplification of weak coherent fields exploiting\\ energy fluctuations of the field}
\date{4 December 2012}
\author{Mikko Partanen}
\author{Teppo H\"ayrynen}
\author{Jani Oksanen}
\author{Jukka Tulkki}
\affiliation{Department of Biomedical Engineering and Computational Science, Aalto University, P.O. Box 12200, 00076 Aalto, Finland}

\begin{abstract}
Quantum optics dictates that amplification of a pure state by any linear deterministic amplifier always
introduces noise in the signal and results in a mixed output state. However, it has recently been shown
that noiseless amplification becomes possible if the requirement of a deterministic operation is relaxed.
Here we propose and analyze a noiseless amplification scheme where the energy required to amplify the
signal originates from the stochastic fluctuations in the field itself. In contrast to previous amplification
setups, our setup shows that a signal can be amplified even if no energy is added to the signal from external
sources. We investigate the relation between the amplification and its success rate as well as the
statistics of the output states after successful and failed amplification processes. Furthermore, we
also optimize the setup to find the maximum success rates in terms of the reflectivities of the beam
splitters used in the setup and discuss the relation of our setup with the previous setups.
\end{abstract}

\maketitle

\section{Introduction}
Any conventional amplification process unavoidably introduces quantum noise into the signal \cite{Caves1982}.
However, this can be circumvented by implementing the amplification nondeterministically so that an amplified
noiseless output signal occurs randomly \cite{Zavatta2011,Ferreyrol2010,Ferreyrol2011,Barbieri2011}.
Noiseless amplification schemes generally rely on applying quantum operations, such as a sequence of
single-photon addition and subtraction, to the optical field \cite{Marek2010,Hayrynen2009,Hayrynen2011}.
In an experimental implementation, the addition and subtraction of photons is typically performed by using single-photon
light sources, beam splitters, and photodetectors \cite{Zavatta2011,Ferreyrol2010,Ferreyrol2011,Barbieri2011},
thus leading to nondeterministic amplification. The noiseless high-fidelity amplification schemes
are expected to become an essential tool for quantum communications and metrology, by recovering
information transmitted over lossy channels or by enhancing the discrimination between partially
overlapping quantum states \cite{Zavatta2011,Josse2006}.

In this paper, we propose and analyze a noiseless amplification scheme where, in contrast to previous works
\cite{Ferreyrol2010,Zavatta2011,Ferreyrol2011}, the energy required to amplify the signal does not originate
from an external energy source (i.e., a single-photon source) but from the stochastic fluctuations in the
field itself. More concretely, the signal is amplified even if no external energy is added to it.
The proposed scheme consists of devices that are frequently used in experiments. Furthermore,
as one of the key challenges in nondeterministic quantum optical amplification of light is the small
success rate, we also consider improving the success probability. We start
by a short summary of the basic principles of noiseless amplification and the related figures of merit.
This is followed by describing the proposed amplification setup and calculations.

\section{Amplification scheme}
\begin{figure}
\includegraphics[width=\columnwidth]{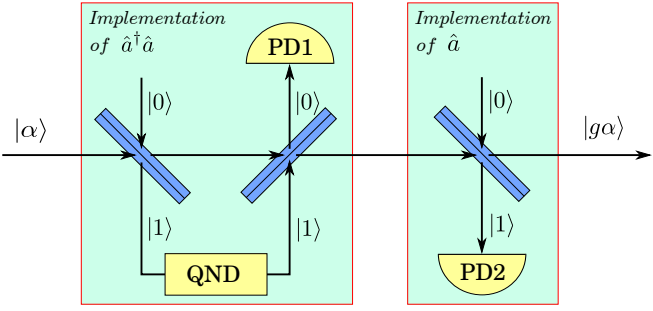}
\caption{\label{fig:scheme} (Color online) Schematic illustration of the noiseless amplification of a weak coherent field. First,
a single photon is subtracted from the field, then added back to the field, and finally again subtracted from the field.
This sequence is also described by the operator $\hat a\hat a^\dag\hat a$.}
\end{figure}

The proposed amplification scheme, illustrated in Fig.~\ref{fig:scheme}, utilizes the energy fluctuations
of the initial field to replace the single-photon source that would otherwise be needed as in the scheme suggested
by Zavatta \emph{et al.} \cite{Zavatta2011}. In our scheme, a similar action is obtained by a configuration where the
successful subtraction of a single photon from the initial field by a beam splitter is verified by a quantum
nondemolition (QND) measurement \cite{Munro2005,Nogues1999,Guerlin2007,Grangier1998,Brune1990,Milburn1984}, which is
followed by adding the photon back to the field by the second beam splitter if no photons are detected at photodetector
PD1. Finally, another photon is subtracted from the field at the third beam splitter, if photodetector PD2 detects a photon.
The final output state resulting from these events is an amplified coherent state with high fidelity when the
QND, PD1, and PD2 detectors detect 1, 0, and 1 photons, respectively.

The action of an ideal noiseless amplifier for coherent states can be described as $|\alpha\rangle\rightarrow|g\alpha\rangle$,
where $|\alpha\rangle$ is the initial coherent field, $|g\alpha\rangle$ is the amplified field, and $g$ is the
gain of amplification. This operation cannot be implemented by deterministic amplifiers, but it can be
approximated probabilistically. It has been shown that the operator $\hat G=\hat a\hat a^\dag$, where
$\hat a$ and $\hat a^\dag$ are the annihilation and creation operators of the field, approximates the
action of amplification for weak coherent fields with nominal gain $g=2$ \cite{Fiurasek2009}. 
The scheme suggested by Zavatta \emph{et al.} \cite{Zavatta2011} is based on this.
The same outcome is also obtained by an operator $\hat G^{\,\prime}=\hat a\hat a^\dag\hat a$
implemented by the setup used in this paper because the input field $|\alpha\rangle$ is an eigenstate of $\hat a$.

\subsection{Output field of the amplifier}
The output fields of our setup have been calculated using the standard Wigner function formalism.
The Wigner function of the initial coherent field $|\alpha\rangle$ is \cite{Schleich2001}
\begin{equation}
W_{\mathrm{coh}}(x,p)=\frac{1}{\pi\hbar}\exp\!\Big[-(\kappa x-\sqrt{2}\mathrm{Re}\alpha)^2-\left(\frac{p}{\hbar\kappa}-\sqrt{2}\mathrm{Im}\alpha\right)^2\Big],
\label{eq:W_coherent}
\end{equation}
where $x$ and $p$ are position and momentum quadratures of the field, $\alpha$ is a complex variable defining the
coherent field amplitude, $\kappa$ is the spring constant of the field oscillator, and $\hbar$ is
the reduced Planck constant. When plotting the Wigner functions, it is conventional to set $\hbar=\kappa=1$
\cite{Schleich2001}.

The entangled Wigner function $W_\mathrm{BS}$ emerging as a result from fields $W_\mathrm{field 1}$
and $W_\mathrm{field 2}$ interfering on a beam splitter is given by \cite{Leonhardt1997,Leonhardt2003}
\begin{align}
W_\mathrm{BS}(x_1,p_1,x_2,p_2) =\; & W_\mathrm{field 1}(tx_1+rx_2,tp_1+rp_2)\nonumber \\
                   &   \times W_\mathrm{field 2}(tx_2-rx_1,tp_2-rp_1),
\label{eq:bs}
\end{align}
where $x_1$, $p_1$, $x_2$, and $p_2$ are the position and momentum quadratures of the transmitted and reflected fields
and the beam splitter reflection and transmission coefficients $r$ and $t$ obey the relation $r^2+t^2=1$.
In our notation $W_\mathrm{field 1}$ is the field incident to the beam splitter from the left and transmitted
field quadratures refer to the field emerging from the beam splitter to the right in Fig.~\ref{fig:scheme}.

The probability of detecting $n$ photons on the reflected field (the field that emerges from the beam splitter
and travels vertically in Fig.~\ref{fig:scheme}) can be expressed as
\begin{align}
P(n)  =\; & 2\pi\hbar\int W_\mathrm{BS}(x_1,p_1,x_2,p_2)\nonumber \\
        & \times W_\mathrm{n}(x_2,p_2)\,dx_1\,dp_1\,dx_2\,dp_2,
\label{eq:probability}
\end{align}
where $W_n$ is the Wigner function of the $n$-photon Fock state $|n\rangle$, to which the reflected field collapses
after the detection, and is expressed as \cite{Schleich2001}
\begin{align}
W_n(x,p) = \; & \frac{(-1)^n}{\pi\hbar}\exp\!\Big[-(\kappa x)^2-\left(\frac{p}{\hbar\kappa}\right)^2\Big]\nonumber\\
              & \times L_n\Big[2(\kappa x)^2+2\left(\frac{p}{\hbar\kappa}\right)^2\Big],
\label{eq:W_fock}
\end{align}
where $L_n(x)$ denotes a Laguerre polynomial of degree $n$. After detecting $n$ photons on the reflected field,
the transmitted field collapses to
\begin{align}
W_\mathrm{T}(x_1,p_1) = \; & \frac{2\pi\hbar}{P(n)}\int W_\mathrm{BS}(x_1,p_1,x_2,p_2)\nonumber \\
                         & \times W_\mathrm{n}(x_2,p_2)\,dx_2\,dp_2.
\label{eq:transmitted}
\end{align}
The collapsed transmitted field is then used as the input for the second beam splitter. Despite the physical
difference between the QND and PD, their effect on the transmitted field is exactly the same, and to calculate
the final output of the setup Eqs.~(\ref{eq:bs})--(\ref{eq:transmitted}) are applied for the remaining two beam
splitters as described in more detail below. For simplicity, we have made the usual assumption that the
photodetectors PD1 and PD2 are ideal. The same assumption is also made for the QND since measurements made
with QND detectors have been reported to yield single-photon Fock states with good accuracy
\cite{Munro2005,Nogues1999,Guerlin2007,Grangier1998,Brune1990,Milburn1984}.

The details of the analysis of how the state is propagated through the setup resulting in the conditional
state of interest are as follows. In the first beam splitter, the initial coherent field $|\alpha\rangle$
in Eq.~\eqref{eq:W_coherent} is mixed with a vacuum state $|0\rangle$ [a zero-photon Fock state $n=0$
in Eq.~\eqref{eq:W_fock}] using Eq.~\eqref{eq:bs}. Then, one photon is measured by the QND detector.
The probability for this and the transmitted field are given by Eqs.~\eqref{eq:probability} and
\eqref{eq:transmitted} with $n=1$. In the second beam splitter, the transmitted field is mixed with
a single-photon Fock state using Eq.~\eqref{eq:bs} since a photon coming from the QND detector is
added to the field. No photons are measured by photodetector PD1. The probability for this and the
transmitted field are given by Eqs.~\eqref{eq:probability} and \eqref{eq:transmitted} with $n=0$.
In the third beam splitter, a photon is subtracted from the field. This is performed by mixing the
field with a vacuum state using Eq.~\eqref{eq:bs} and using Eqs.~\eqref{eq:probability} and
\eqref{eq:transmitted} with $n=1$ for calculating the probability and the transmitted field that
is the final output state of the setup. The total probability for this successfully amplified output
state $P_\mathrm{succ}$ is the product of the mentioned three photon detection probabilities given by
\begin{equation}
P_\mathrm{succ}=(1+|t_1t_2t_3\alpha|^2(3+|t_1t_2t_3\alpha|^2))
|r_1r_2r_3\alpha|^2 e^{|t_1t_2t_3\alpha|^2-|\alpha|^2},
\label{eq:optfunc}
\end{equation}
where $r_i$ and $t_i$, $i=1,2,3$, are reflectivities and transmittivities of the beam splitters in the setup
obeying $r_i^2+t_i^2=1$.

\subsection{Effective gain and fidelity of the amplified state}
In the calculations depending on the parameters of the setup, effective gain values different from the
nominal gain of 2 can be found. The effective gain can be defined as the ratio of the expectation
values of the annihilation operator $\hat a$ for the output and input fields \cite{Zavatta2011}
\begin{equation}
g_\mathrm{eff}=\frac{|\langle\hat a_\mathrm{out}\rangle|}{|\langle\hat a_\mathrm{in}\rangle|},
\label{eq:geff}
\end{equation}
which corresponds to the effective amplification of the electric-field amplitude.
In the Wigner function formalism, the expectation value of the annihilation operator can be calculated using
the operator correspondence relation as \cite{Gardiner2004}
\begin{align}
 \langle\hat a\rangle =\; & \int\left[\frac{\kappa}{\sqrt{2}}\left(x+\frac{i\hbar}{2}\frac{\partial}{\partial p}\right)
 +\frac{i}{\sqrt{2}\hbar\kappa}\left(p-\frac{i\hbar}{2}\frac{\partial}{\partial x}\right)\right]\nonumber\\
                        & \times W(x,p)\,dx\,dp.
 \label{eq:annih}
\end{align}
The calculations produce the following expression for the effective gain:
\begin{equation}
g_\mathrm{eff}=\frac{t_1 t_2 t_3 (2+4|t_1 t_2 t_3\alpha|^2+|t_1 t_2 t_3\alpha|^4)}
{1+3|t_1 t_2 t_3\alpha|^2+|t_1 t_2 t_3\alpha|^4}.
\label{eq:gefffull}
\end{equation}

It is also useful to quantify how much the output state differs from an ideally amplified coherent state.
A practical measure for this purpose is the fidelity $F$, which is the overlap between
the states calculated using Wigner functions $W_1$ and $W_2$ of the compared fields \cite{Lee2000}
\begin{equation}
F(W_1,W_2)=2\pi\hbar\int W_1(x,p)W_2(x,p)\,dx\,dp.
\label{eq:fidelityw}
\end{equation}
The fidelity obtained for the successfully amplified field with respect to a coherent
field $|g_\mathrm{eff}\alpha\rangle$ is
\begin{equation}
F_\mathrm{eff}=\frac{(1+2g_\mathrm{eff}t_1t_2t_3|\alpha|^2+g_\mathrm{eff}^2t_1^2t_2^2t_3^2|\alpha|^4)
 e^{-(g_\mathrm{eff}^2-t_1t_2t_3)^2|\alpha|^2}}
{1+3|t_1 t_2 t_3\alpha|^2+|t_1 t_2 t_3\alpha|^4}.
\label{eq:fidelityeff}
\end{equation}

\begin{figure}
\includegraphics[width=\columnwidth]{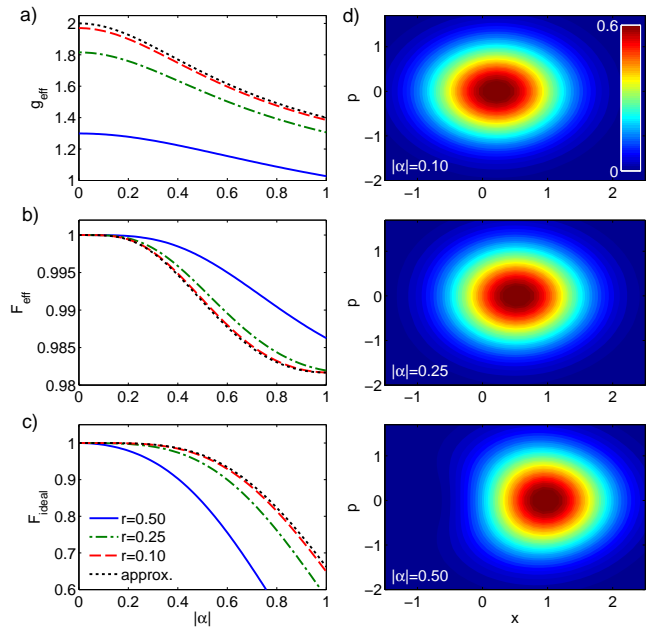}
\caption{\label{fig:succ} (Color online) (a) Effective gain as a function of the input field amplitude for three different
beam splitter reflectivities and for the analytic low reflectivity approximation used by Zavatta \emph{et al.}
\cite{Zavatta2011}. (b) The effective fidelity calculated with respect to a coherent field $|g_\mathrm{eff}\alpha\rangle$.
(c) The fidelity calculated with respect to an ideally amplified field $|2\alpha\rangle$ for comparison
with the results obtained by Zavatta \emph{et al.} \cite{Zavatta2011}. (d) The contour plots of the Wigner functions
for three amplified coherent fields with different input amplitudes when the beam splitter reflectivity is $r=0.4$.}
\end{figure}

The effective gain, fidelity, and Wigner functions of successfully amplified fields are presented in
Fig.~\ref{fig:succ}. In Fig.~\ref{fig:succ}(a) the effective gain is very close to the nominal
gain value $g=2$ for small values of $|\alpha|$ and $r$. As the input field
amplitude or the beam splitter reflectivity increases, the gain decreases. Increasing the input field
amplitude results in the reduction of the effective fidelity $F_\mathrm{eff}$ as shown in Fig.~\ref{fig:succ}(b),
where the fidelity is calculated with respect to a coherent field $|g_\mathrm{eff}\alpha\rangle$.
However, the reduction of fidelity can be partly compensated by increasing the beam splitter reflectivity.
Figure \ref{fig:succ}(c) shows the fidelity calculated with respect to an ideal maximally amplified coherent
field $|2\alpha\rangle$ for comparison with the results obtained by Zavatta \emph{et al.} \cite{Zavatta2011}
for the setup, including a specific single-photon source. The values for this ideal fidelity $F_\mathrm{ideal}$
decrease faster than the effective fidelities $F_\mathrm{eff}$ in Fig. \ref{fig:succ}(b) due to the
reduction in $g_\mathrm{eff}$ for stronger input fields. Thus $F_\mathrm{eff}$ is a better measure for the
quality of the resulting output field. One can also see that $F_\mathrm{ideal}$ decreases when the beam splitter
reflectivity increases while the opposite is true for $F_\mathrm{eff}$. This is also due to the reduction
in the effective gain. The contour plots in Fig.~\ref{fig:succ}(d) demonstrate how the Wigner function
deforms in the amplification. For small initial field amplitudes, the output field is very close
to a pure coherent field, but it increasingly deviates from a coherent state when the initial
field amplitude increases.

\subsection{Optimizing the scheme}
Next we investigate how to optimize the probability of successful amplification while maintaining a given
effective gain. The optimization problem for maximizing the probability of successful amplification
[Eq.~\eqref{eq:optfunc}] with a constraint requiring the effective gain [Eq.~\eqref{eq:gefffull}]
exceeding a threshold value $g_\mathrm{eff,0}$ can be presented as
\begin{equation}
\max_{g_\mathrm{eff}\ge g_\mathrm{eff,0}} P_\mathrm{succ}(|\alpha|,r_1,r_2,r_3).
\label{eq:optproblem}
\end{equation}
Here, $|\alpha|$ is the input field amplitude and the beam splitter reflectivities are $r_1$, $r_2$, and $r_3$.
The four-dimensional nonlinear optimization problem in Eq.~\eqref{eq:optproblem} was solved using a barrier function method
\cite{Bazaraa2006}. For a certain $g_\mathrm{eff,0}$, one finds a single maximum $P_\mathrm{opt}$ with an optimal
input field amplitude $|\alpha|_\mathrm{opt}$ and beam splitter reflectivities $r_\mathrm{1,opt}$, $r_\mathrm{2,opt}$,
and $r_\mathrm{3,opt}$. The optimization problem was solved multiple times changing the minimum effective gain
parameter $g_\mathrm{eff,0}$. It was found that at the optimum all the beam splitter reflectivities have the
same value $r_\mathrm{i,opt}=r_\mathrm{opt}$, $i=1,2,3$.

\begin{figure}
\includegraphics[width=\columnwidth]{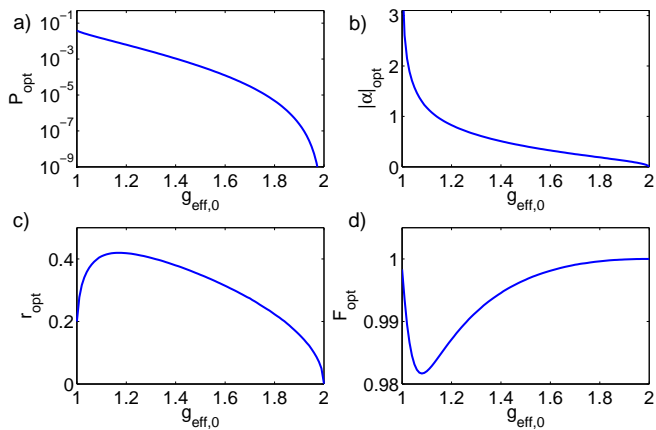}
\vspace{-0.5cm}
\caption{\label{fig:opti}(Color online) Probability of successful amplification was maximized in an effective gain constrained
optimization problem. The optimal (a) probability of successful amplification $P_\mathrm{opt}$, (b) the input field
amplitude $|\alpha|_\mathrm{opt}$, (c) the beam splitter reflectivity $r_\mathrm{opt}$, and (d) the output field
fidelity $F_\mathrm{opt}$ are plotted as a function of the required minimum effective gain
parameter $g_\mathrm{eff,0}$.}
\end{figure}

Figure \ref{fig:opti} shows how the optimized
probability of successful amplification $P_\mathrm{opt}$, the input field amplitude $|\alpha|_\mathrm{opt}$,
the beam splitter reflectivity $r_\mathrm{opt}$, and the fidelity of the successfully amplified state
$F_\mathrm{opt}$ evolve as a function of the minimum effective gain parameter $g_\mathrm{eff,0}$.
The probability of successful amplification in Fig.~\ref{fig:opti}(a) decreases exponentially
when the effective gain increases. For instance, if one wants to have an effective gain of 1.4,
the maximum success probability of $10^{-3}$ is achievable with $|\alpha|_\mathrm{opt}=0.51$ and
$r_\mathrm{opt}=0.38$. For comparison, Ferreyrol \emph{et al.} \cite{Ferreyrol2010,Ferreyrol2011} reported
success rates of order $10^{-2}$ for a conventional scheme based on quantum scissors \cite{Ralph2008}.
However, their scheme required a single photon source whose effect is not included in the reported
success rates. Thus the obtained values can not be directly compared.
In Fig.~\ref{fig:opti}(b) one sees that for useful values of $g_\mathrm{eff,0}$ the optimal input field
amplitude is limited to $|\alpha|_\mathrm{opt}<1$. The optimal beam splitter reflectivity in Fig.~\ref{fig:opti}(c)
has a maximum $r_\mathrm{opt}=0.42$ at the effective gain $g_\mathrm{eff,0}=1.18$ and it approaches zero when the
required effective gain approaches 2. The fidelity curve in Fig.~\ref{fig:opti}(d) has a minimum $F_\mathrm{opt}=0.982$
at $g_\mathrm{eff,0}=1.08$. In the nominal gain limit, the fidelity approaches unity.

\subsection{Failed amplification}
So far, we have only discussed the case of successful amplification. For completeness, we next analyze the other
possible output states. If the initial field is weak and the beam splitter reflectivities are $<\hspace{-2.5pt}0.5$, the probability
that any of the photodetectors detects more than one photon is typically $<\hspace{-4pt}10^{-2}$. This probability is not completely
negligible but since it is still small and the occurrence of these processes can be detected, we can focus on the
processes where only at most one photon is detected at a time. These single photon processes and the corresponding
eight possible output states are described in Table \ref{tbl:states}. The output states can be experimentally
identified by photon detection measurement outcomes. The first state is the successfully amplified field
and the last row shows the probability that more than one photon is detected by some of the photodetectors.

The fidelities in Table \ref{tbl:states} clearly show that states from 5 to 8 are exactly coherent.
This can be understood by considering what happens if the first photon subtraction fails. In this case, the
output from the first beam splitter can be shown to be $|t\alpha\rangle$, which is a perfectly coherent field.
This further means that at beam splitters 2 and 3, only single-photon subtraction or no photon subtraction
can take place. Both operations result in coherent fields, albeit with reduced amplitude. This is because
the photon subtractions only decrease the amplitude and keep the state coherent since coherent states
are eigenstates of the annihilation operator \cite{Hayrynen2009,Kim2008a}. The states 3 and 4 are not
exactly coherent since, in these cases, the input field arriving to the second beam splitter from the
QND device is a single-photon Fock state producing superposition states at the output.

\begin{table}
\caption{\label{tbl:states}Photon detection measurement outcomes, amplitude expectation values $|\langle\hat a\rangle|$,
degradations of fidelities $1-F_\mathrm{eff}$, and probabilities $P$ for possible single-photon process output states
of the amplification setup with $g_\mathrm{eff}=1.4$. The initial field is a coherent field with $|\alpha|=0.5$ and the reflectivity of
the beam splitters is $r=0.4$. Successful amplification corresponds to the first state.
}
\centering
\renewcommand{\arraystretch}{1.5}
\begin{tabular}{ccccccc}
\hline\hline
        & \multicolumn{3}{c}{Measurements} & & & \\
\cline{2-4}
  State & QND & PD1 & PD2 & $|\langle\hat a\rangle|$ & $1-F_\mathrm{eff}$ & $P$\\
\hline
 1 & 1 & 0 & 1 & 0.686 & $4.84\times 10^{-3}$  & $1.36\times 10^{-3}$\\
 2 & 1 & 0 & 0 & 0.720 & 0.362                & $5.58\times 10^{-3}$\\
 3 & 1 & 1 & 1 & 0.292 & $3.79\times 10^{-5}$  & $5.27\times 10^{-4}$\\
 4 & 1 & 1 & 0 & 0.310 & $1.60\times 10^{-5}$  & $2.88\times 10^{-2}$\\
 5 & 0 & 1 & 1 & 0.385 & 0                    & $8.57\times 10^{-4}$\\
 6 & 0 & 1 & 0 & 0.385 & 0                    & $3.03\times 10^{-2}$\\
 7 & 0 & 0 & 1 & 0.385 & 0                    & $2.55\times 10^{-2}$\\
 8 & 0 & 0 & 0 & 0.385 & 0                    & 0.903\\
 other & & & & & & $3.84\times 10^{-3}$\\
\hline\hline
\end{tabular}
\end{table}

The amplitude expectation values in Table \ref{tbl:states} showed that the amplitude of the coherent output
states 5--8 $|\langle\hat a\rangle|=0.385$ is clearly smaller than the amplitude of the initial field $|\alpha|=0.5$.
This follows from the relatively large reflectivity of $r=0.4$. For a smaller reflectivity of $r=0.1$, the amplitude
expectation value for the exactly coherent output states is $|\langle\hat a\rangle|=0.493$, which is much closer
to the initial field amplitude. Since this output is also the most probable output and, in the case of small
reflectivities, it is a nearly unchanged coherent state one could also try to repeat the amplification
process in order to increase the probability of successful amplification. However, experimental realization
of the repeated amplification setup would be challenging.

\section{Conclusions}
In conclusion, we have studied noiseless amplification of coherent signals in a setup where all the energy added
to the amplified signal originates from the fluctuations in the quantum field in a purely stochastic manner, i.e.,
the field is amplified even when no additional energy is added to the field from external sources in contrast to
the previously reported noiseless amplifiers. We have shown that the probability of successful amplification can
be maximized by finding optimal values for the beam splitter reflectivities depending on the desired effective
gain. Our results show that the purely stochastic amplification scheme can amplify weak coherent fields with
very good fidelities much like the conventional stochastic amplification setups relying on single-photon sources.
Most importantly, however, all parts of our setup have been experimentally demonstrated so that the proposed
amplification scheme is experimentally feasible and may allow experimentally demonstrating noiseless amplification
that requires no external energy.

\end{document}